# Metamorphic InAs$_{1-x}$Sb$_x$/InAs$_{1-y}$Sb$_y$ superlattices with ultra-low bandgap as a Dirac material.

Sergey Suchalkin[1], Gregory Belenky[2], Maksim Ermolaev[1], Seongphill Moon[2,3], Yuxuan Jiang[2,3a], David Graf[2], Dmitry Smirnov[2], Boris Laikhtman[4], Leon Shterengas[1], Gela Kipshidze[1], Stefan P. Svensson[5], Wendy L. Sarney[5].

[1]*State University of New York at Stony Brook, Stony Brook, New York, 11794-2350, USA*

[2]*National High Magnetic Field Laboratory, Tallahassee, FL 32310, USA*

[3]*Department of Physics, Florida State University, Tallahassee, Florida 32306, USA*

[3a]*School of Physics, Georgia Institute of Technology, Atlanta, Georgia 30332, USA*

[4]*Racah Institute of Physics, Hebrew University, Jerusalem 91904, Israel*

[5]*U.S.Army Research Laboratory, 2800 Powder Mill Rd, Adelphi, Maryland 20783, USA*

It was experimentally demonstrated that short-period metamorphic InAs$_{1-x}$Sb$_x$/InAs$_{1-y}$Sb$_y$ superlattices with ultra low bandgap have properties of a Dirac material. Cyclotron resonance and interband magneto-absorption peaks in superlattices with ultra-low bandgaps demonstrate a square root dependence on the magnetic field for a range up to 16 T (energy range up to 300meV). This directly indicates the linearity of the electron dispersion. Experimental estimates of the Fermi velocity gives $v_F = 7 \cdot 10^5 \frac{m}{s}$. The Fermi velocity can be controlled by varying the overlap between electron and hole states in the superlattice. The dependence of the cyclotron resonance energy on the magnetic field parallel to the superlattice plane demonstrates that the electron dispersion in the growth direction can be characterized by an effective mass of 0.028m$_0$ in a superlattice with a period of 6 nm and 0.045m$_0$ in a superlattice with a period of 7.5 nm. Extreme design flexibility makes the short-period metamorphic InAs$_{1-x}$Sb$_x$/InAs$_{1-y}$Sb$_y$ superlattice a new prospective platform for studying the effects of charge carrier chirality and topologically nontrivial states in structures with the inverted bandgaps.

**PACS:**

Several materials systems have been found with specific crystal symmetries that produce energy dispersions that deviate from the common quadratic dependence.[1-3]. The Dirac fermion is an example of a solid state excitation with a dispersion similar to that of a free relativistic particle. In contrast to conventional semiconductors, the dispersions of the electrons and holes in Dirac materials are characterized by the same effective mass, which is directly related to the spectral gap [4]. Dirac fermions were demonstrated in various materials and systems such as d-wave superconductors [3], graphene [see [3] and references therein], and narrow-gap semiconductors such as HgCdTe [5-6], Bi$_2$Te$_3$ [7], ZrTe$_5$[8]. As the bandgap of a semiconductor material is tuned to zero by alloy composition [5], temperature [9], strain [10-11] or by dimensional quantization in semiconductors with an inverted bandgap [6], strong intermixing of electron and hole states leads to the Dirac-type dispersion.

In type II semiconductor heterostructures with a "broken gap" band alignment, the effective bandgap is determined by dimension quantization and can be sufficiently different from the bulk bandgaps of the SL layer materials. Type II InAs/GaSb composite quantum wells with zero and inverted band gaps demonstrate a rich phase diagram containing band insulator and quantum spin Hall insulator states [12-13]. The ordered InAsSb alloy has recently attracted attention as a new potential platform for observing topologically nontrivial phases and realizing Majorana zero modes [14-15].

The electronic structure of a type II SL with zero effective bandgap is, generally, different from that of a gapless bulk material since the electrons and holes in the type II SL occupy different layers. If the typical SL bandgap width $E_g^{SL}$ is much larger than the energy scale associated with the in-plane motion $E_\parallel$,

$$E_g^{SL} \gg E_\parallel \quad (1),$$

the latter can be considered as a perturbation[16], and the in-plane energy spectrum is similar to that of the gapless bulk material. Condition (1) can also be expressed as:

$$k_\parallel \ll \frac{\pi}{d} \quad (2),$$

where $k_\parallel$ is the in-plane wave vector and $d$ is the SL period[16].

It is possible to design metamorphic InAs$_{1-x}$Sb$_x$/InAs$_{1-y}$Sb$_y$ superlattices [14, 17] that reach nearly zero effective bandgap for a SL period d~6-7 nm. The short period is made possible by the relative positions and the small (<0.2 eV) bulk bandgaps of the constituents of the SL. In addition, the virtual substrate approach [18] used for the SL fabrication allows strain variation within the SL period. Strain of opposite signs applied to electron- and hole containing layers can further reduce the bandgap.

In this paper, we demonstrate experimentally that Dirac-type carrier dispersion can be realized in a III-V semiconductor system using a type II superlattice with an effective bandgap close to zero. Interband magneto-absorption and cyclotron resonance were used to probe the electron dispersion for both the in-plane (Faraday geometry) and the growth (Voigt geometry) direction.

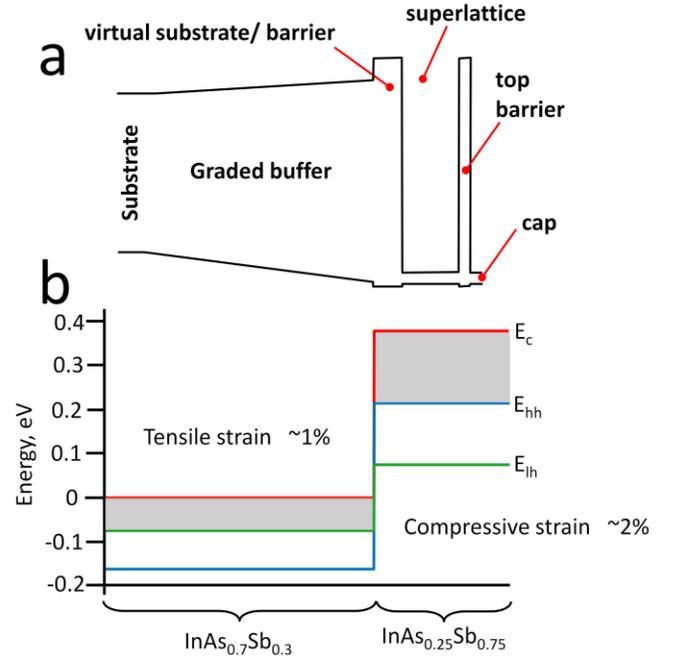

FIG. 1. Schematic band diagram of the overall structure (a) and bulk bandgap energies of the SL materials (b)

A schematic band diagram of the structure is presented in Figure 1. The growth was done with solid-state molecular beam epitaxy on unintentionally doped (100) GaSb substrates. A GaSb buffer layer e was followed by a 2800 nm graded buffer with the composition varied from GaSb to Al$_{0.58}$In$_{0.42}$Sb. This was terminated with a wide bandgap Al$_{0.58}$In$_{0.42}$Sb virtual substrate/lower barrier with a lattice constant of 6.25 A [18]. The thickness of the absorber layer, consisting of a InAs$_{0.7}$Sb$_{0.3}$/InAs$_{0.25}$Sb$_{0.75}$ SL, is 1μm. The InAs$_{0.7}$Sb$_{0.3}$ layers are under 1% tensile strain while the InAs$_{0.25}$Sb$_{0.75}$ layers are under

~2% compressive strain. Strain balance was achieved by using a layer thickness ratio of 2. Two samples with 6 nm (structure 1) and 7.5 nm (structure 2) periods were made. In both samples, the absorber layer was terminated with a 200 nm $Al_{0.58}In_{0.42}Sb$ barrier and a 50 nm cap made of the same SL. To avoid formation of 2D electron pockets at the sample boundaries, the barriers and the cap were p-doped at a concentration of $10^{16}$ cm$^{-3}$. To measure magneto-absorption the samples were placed in a liquid He cryostat with a superconductive magnet.

The magnetic field dependence of the cyclotron resonance energy measured in the Faraday geometry is presented in Figure 2.

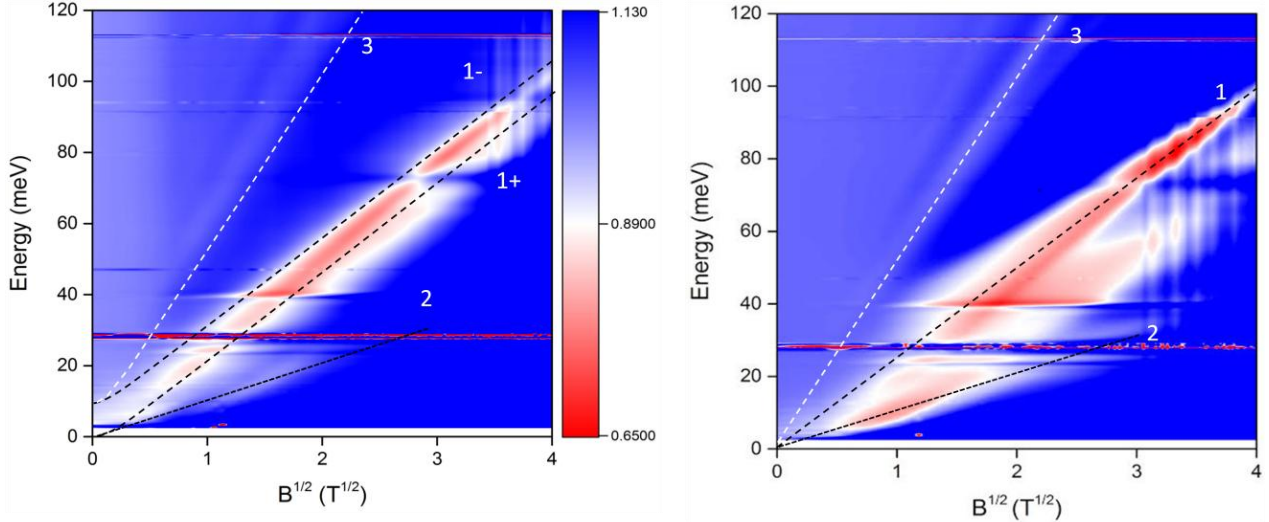

FIG. 2 Color plot of the relative transmission T(B)/T(0) as a function of energy and the square root of the magnetic field for the SL with a 6 nm period (a) and the SL with a 7.5 nm period (b). Lines 1+, 1- and 2 represent the transitions between electron LLs; line 3 corresponds to the interband transition.

The dependence of the cyclotron resonance (CR) peak energy on $\sqrt{B}$ is linear in the entire range of the magnetic field. There is a ~ 10meV splitting of the CR line (Figure 3) which is not clearly seen in the color plot. We attributed the CR lines $1^+$ and $1^-$ in Figure 2a to the cyclotron resonance (CR) transitions between the $0^+$ and $0^-$ ground and the first electron Landau levels (LLs)[6]. The LL energies of a massive Dirac fermion can be approximated as[16]:

$$E_n \approx \frac{E_g}{2} \pm \sqrt{\frac{E_g^2}{4} + n\frac{2\hbar^2 v_F^2}{l_B^2}} \quad (3)$$

Here $v_F$ is the Fermi velocity, $E_g$ for structure 1 is taken as 10 meV, $E_g$ for structure 2 is zero, $l_B = \sqrt{\frac{\hbar c}{eB}}$ is the magnetic length and the +/- sign corresponds to the electrons and holes (Appendix A). The best fit of the experimental dependencies of CR energy on $\sqrt{B}$ is obtained at $v_F = 7 \cdot 10^5 \frac{m}{s}$ for both structures (Figure 2). The splitting between the 1+ and 1- transitions, which is equal to the SL bandgap [6], is not clearly seen in the experimental data since the bandgap is comparable with the CR line width. Line 2 is the calculated transition energy from electron LL 1 to 2). This transition is clearly observed in sample 2 at magnetic fields below ~ 2T. At higher magnetic fields, the LL2 is depleted due to higher LL degeneracy and the CR transition occurs between the 0$^{th}$ and 1$^{st}$ electron LLs. The dashed line 3 in Figure 2 is the calculated energy of the transition between the first LLs of electrons and holes as a function of the square root of the magnetic field. The transition energy is calculated using the assumption that electrons and holes have the same linear dispersion characterized by $v_F = 7 \cdot 10^5 \frac{m}{s}$. Experimental CR and electron-hole transition energies precisely follow a $\sqrt{B}$ dependence for both samples. The results are qualitatively similar to those observed in nearly gapless HgCdTe [5-6]. Line 2 in Figure 2a,b is the transition between the first electron and the first hole Landau levels. The energy of this transition is ~ 2 times higher than the CR energy, which is an indication that the electrons and holes have the same Landau quantization energies and similar energy dispersions.

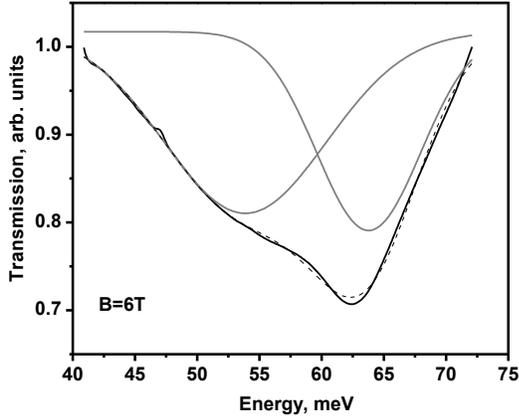

FIG. 3. CR absorption peak of the SL with the period 6nm. Solid line – experiment, dashed line –two peak Gaussian fitting.

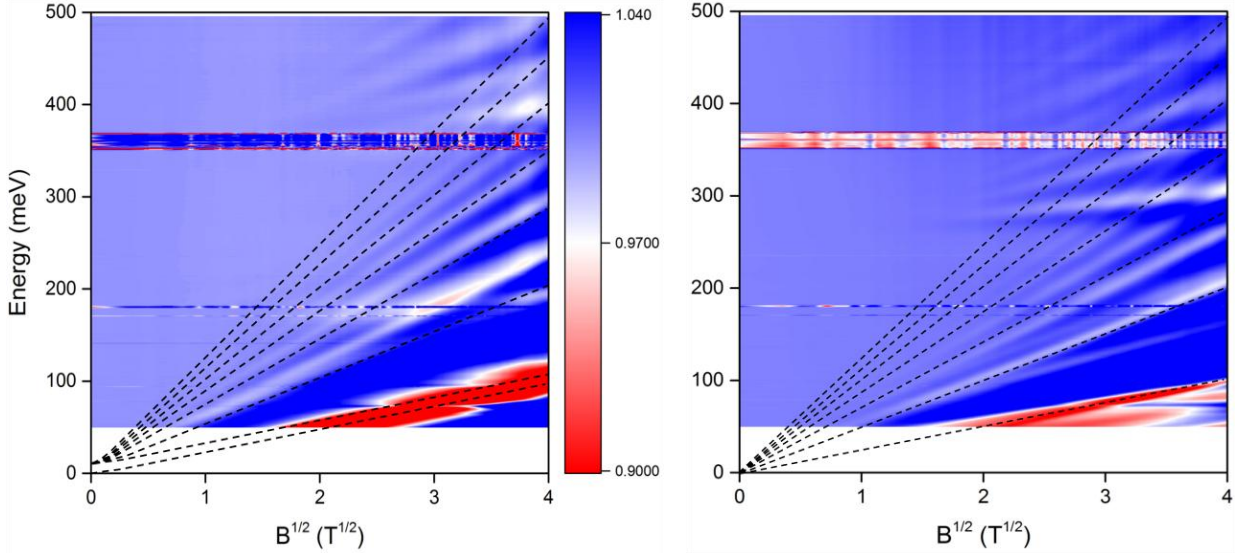

FIG. 4 Color plot of relative transmission T(B)/T(0) as function of energy and square root of magnetic field for structure 1 with d=6nm (a) and structure 2, d=7.5nm (b). The dashed lines is the transition energies calculated according to expression 3.

A color plot of the mid-IR magneto-absorption is shown in Figure 4. The dashed lines are the transition energies, calculated using the expression (3). A possible reason for deviation of the observed transition energies from the theoretical ones is the asymmetry between electron and hole spectra at higher energies. The detailed description of the transition energies vs magnetic field with more sophisticated model will be presented elsewhere. One of the striking features of the magneto-absorption spectra is that the all transition energies demonstrate $\sqrt{B}$ dependence for the entire observed energy range. This dependence is observable at energies that are considerably less than the bandgaps of both ternary alloys of the SL layers (Figure 2). These are ~ 100meV for 1% tensile strained $InAs_{0.7}Sb_{0.3}$ and ~ 200meV for 2% compressively strained $InAs_{0.25}Sb_{0.75}$. This is an indication that the parabolic region in the in-plane energy spectrum of a short period type II superlattice is determined by the effective SL bandgap rather than the bulk bandgaps of the corresponding SL materials. As the effective bandgap of the SL: is close to zero, the in-plane energy spectrum is Dirac-type and consists of two symmetrical branches of electrons and heavy holes. The Fermi velocity $v_F = \frac{|P_{hh,e}|}{\sqrt{2}m_0}$, where $P_{hh,e}$ – averaged over the SL period momentum matrix element between electron and heavy hole subbands [16] is:

$$P_{hh,e} = P_1 \int_0^{d1} \xi_e(z)\xi_{hh}^*(z)dz + P_2 \int_{d1}^{d} \xi_e(z)\xi_{hh}^*(z)dz \quad (4).$$

Here $P_1$, $P_2$ are the momentum matrix elements between the electron and heavy hole subbands in $InAs_{0.7}Sb_{0.3}$ and $InAs_{0.25}Sb_{0.75}$; $d_1$ is the thickness of the $InAs_{0.7}Sb_{0.3}$ layer; and $\xi_e(z), \xi_{hh}(z)$ are Kronig-Penney envelop functions for electrons and heavy holes corresponding to the motion in the growth direction

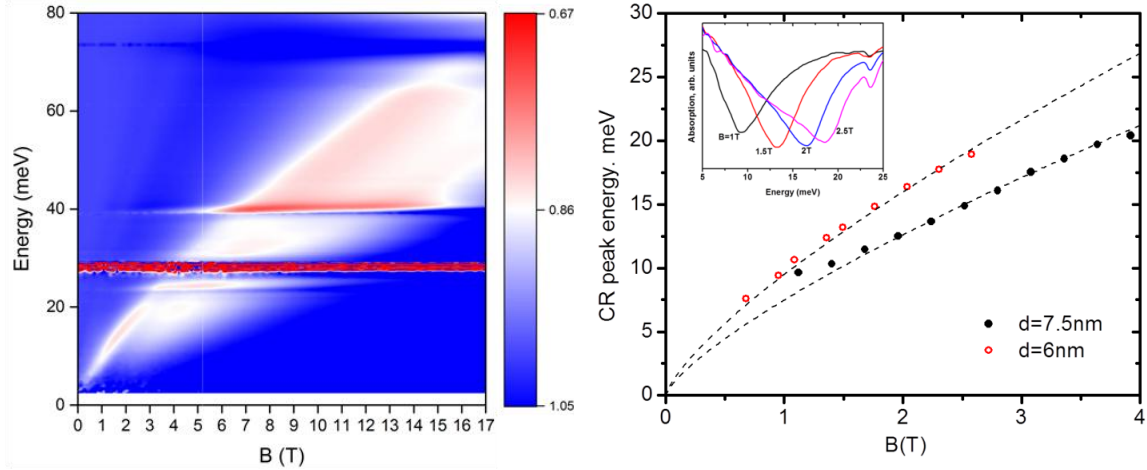

FIG. 5. Color plot of relative transmission T(B)/T(0) as function of energy and magnetic field, parallel to the SL layers, for the SL with a period o f 6 nm (a) and magnetic field dependence of the CR energy (b). Dashed lines are the CR energies calculated using expression (6). Inset: CR lines at different magnetic fields.

The magnitude $2P_1^2/m_0$ is ~ 19eV for InAs and 20.5eV for InSb [19], so $P_1 \approx P_2 \equiv P$ and

$$v_F \approx \frac{P}{\sqrt{2}m_0} I \approx 1.2 \cdot 10^6 \cdot I \left[\frac{m}{s}\right] \quad (5),$$

Where $I \equiv \left|\int_0^d \xi_e(z)\xi_{hh}^*(z)dz\right|$ is the overlap between electron and heavy hole states in the SL. Comparison of theoretical expression for $v_F$ (5) with the experimentally obtained $v_F = 7 \cdot 10^5 \frac{m}{s}$ gives I=0.58. This is a reasonable value considering that the holes are well localized in the $InAs_{0.25}Sb_{0.75}$ layers. Adjusting the design parameters of the metamorphic SLs makes it possible to vary the electron-hole overlap while keeping the effective $E_g$ of the SL close to zero. This gives an effective tool to control the Fermi velocity.

A color plot of the far-IR magneto absorption for the magnetic field, parallel to the SL layer (Voigt geometry) is shown in Figure 4(a). The magnetic field in the Voigt geometry mixes the in-plane and vertical motion of the carriers. It follows from the Kronig-Penney model, that at vertical wave vectors $k_\perp \ll \frac{\pi}{d}$ the carrier dispersion in the SL growth direction is parabolic and can be characterized by an effective mass $m_\perp$. This leads to a sufficiently different dependence of the CR energy on the magnetic field compared with the Faraday geometry. A clear CR peak is observed for magnetic fields below 4 T (Figure 4, inset). The peak broadens at higher magnetic fields. This can be attributed to the onset of interface-bound CR [20-21], as the magnetic length becomes closer to the SL period. At magnetic fields below 4 T the magnetic length exceeds 12.8 nm, which is larger than the SL periods. At these conditions, the CR mass in Voigt geometry can be estimated as a geometric mean of the magnetic field dependent in-plane CR mass and the constant vertical CR mass $m_\perp$. The magnetic field dependence of the CR energy $E_{CR}$ can be estimated as:

$$E_{CR} \approx \left(\frac{\sqrt{2}\hbar^3 v_f}{m_\perp l_B^3}\right)^{1/2} \quad (6)$$

The dashed lines in Figure 4 (b) is the CR peak energy calculated using expression (6) with $m_\perp = 0.028 m_0$ for the SL with d=6 nm and $m_\perp = 0.045 m_0$ for the SL with d=7.5 nm.

The interband transitions are not observable at energies less than 40 meV, probably due to the effect of band filling. The carrier concentration obtained from the hall measurements gives N ~ $10^{16}$ cm$^{-3}$ for the SL with 6 nm period and N~ $1.7 \times 10^{16}$ cm$^{-3}$ for the SL with 7.5 nm period. Assuming that the dispersion along the growth direction is parabolic and characterized by the effective mass $m_\perp$, and the in-plane dispersion is linear and characterized by the Fermi velocity $v_F$, the Fermi energy $E_F$ can be expressed through the carrier density $N$ as:

$$E_F = \left(\frac{15\pi^2 \hbar^3 v_F^2}{4\sqrt{2}\sqrt{m_\perp}} N\right)^{2/5} \quad (7)$$

The estimations of the Fermi energy give 25 meV for the 6 nm SL and 30 meV for the 7.5 nm SL. As one can see from Figure 2(b), the switch from the LL0-LL1 CR transition to LL1-LL2 occurs at ~ CR energy ~ 30 meV. The sum of the Fermi energy and the effective SL bandgap gives the lowest energy at which the interband optical transitions can be observed.

In conclusion, we have demonstrated experimentally that the in-plane dispersion in ultra low bandgap metamorphic $InAs_{1-x}Sb_x/InAs_{1-y}Sb_y$ superlattices is Dirac type with the Fermi velocity $v_F = 7 \cdot 10^5 \frac{m}{s}$. The Fermi velocity is proportional to the overlap of the electron and hole states in the SL, which in our case is ~0.58. It was found that the electron dispersion in the growth direction is parabolic at low energies and can be characterized by an effective mass of $0.028m_0$ in the SL with a period of 6 nm and $0.045m_0$ in the SL with a period of 7.5 nm. Applying the virtual substrate approach makes it possible to control the parameters of the in-plane and vertical carrier electron dispersion while keeping the same value of the effective bandgap.

The work is supported by U.S. Army Research Office Grant W911TA-16-2-0053. S.M., Y.J. and D.S. acknowledge the support from the U.S. Department of Energy (grant number DE-FG02-07ER46451) for IR magneto-spectroscopy measurements that were performed at the National High Magnetic Field Laboratory, which is supported by National Science Foundation (NSF) Cooperative Agreement Number DMR-1157490 and the State of Florida.

Appendix A

In $InAs_{1-x}Sb_x$ compounds the electron effective mass at the bottom of the conduction band is $m^* \approx 0.01 - 0.025 m_0$ [22] and in SL with period around 10 nm quantization energy $\pi^2 \hbar^2 / 2m^* d^2 > 600 meV$. This energy is significantly larger than the in-plane energies in the experiment. Therefore in the calculation of the electron spectrum the Hamiltonian of the system can be broken in two parts,

$$H = H_\perp + H_\parallel \tag{A1}$$

where $H_\parallel$ that describes in-plane dynamics can be considered as a perturbation[16]. In the leading approximation Schrödinger equation

$$H\Psi = E\Psi \tag{A2}$$

is reduced to one dimensional equation

$$H_\perp \Xi(z) = \varepsilon \Xi(z) \tag{A3}$$

The width of SL layers is much larger than the lattice constant and for $H_\perp$ it is possible to use the effective mass approximation. This approximation means diagonalization of $H_\perp$ separately in each SL layer. As a result, equations for electrons, light holes and heavy holes are separated. We neglect spin-orbit split band because of its large separation energy from other types of carriers. We also neglect small band mixing at SL interfaces. Than for each type of carries the problem is reduced to Kronig-Penney model with different effective masses in wells and barriers. Wave functions $\xi_\alpha(z)$ for different carriers ($\alpha = e$ for electrons, $\alpha = hh$ for heavy holes and $\alpha = lh$ for light holes) differ only by the values of constants, the effective mass and height of the barrier, and don't depend on spin. They are characterized by the subband number and quasi-wave vector $q$ in the growth direction. Due to high SL quantization energy we discard excited subbands and include only first subbands for all carriers in further calculation. All $\xi_\alpha(z)$ are components of the same wave functions and are characterized by the same value of $q$ that will be omitted to shorten the notations. All $\xi_\alpha(z)$ are normalized at the SL period.

We take $H_\parallel$ in the approximation corresponding the Kane model where, as well as in the growth direction, we neglect spin-orbit split band. We also neglect valence band anisotropy that makes it possible to obtain analytic results and would give only small corrections to the electron spectrum. Then in the basis

$$\begin{pmatrix}u_+\\u_-\end{pmatrix}, \quad u_+ = \begin{pmatrix}|S,1/2\rangle\\|3/2,3/2\rangle\\|3/2,1/2\rangle\end{pmatrix} \quad u_- = \begin{pmatrix}|S,-1/2\rangle\\|3/2,-3/2\rangle\\|3/2,-1/2\rangle\end{pmatrix} \tag{A4}$$

both $H_\perp$ and $H_\parallel$ in each of the layers are $6\times 6$ matrices. That is Eq.(3) has 6 solutions

$$\Xi_{+\alpha}(z) = \begin{pmatrix}\Xi_{+\alpha}(z)\\0\end{pmatrix}, \quad \Xi_{-\alpha}(z) = \begin{pmatrix}0\\\Xi_{-\alpha}(z)\end{pmatrix} \tag{A5}$$

$$\Xi_{\pm e}(z) = \begin{pmatrix}\xi_e(z)\\0\\0\end{pmatrix}, \quad \Xi_{\pm hh}(z) = \begin{pmatrix}0\\\xi_{hh}(z)\\0\end{pmatrix}, \quad \Xi_{\pm lh}(z) = \begin{pmatrix}0\\0\\\xi_{lh}(z)\end{pmatrix} \tag{A6}$$

and the in-plane Hamiltonian is

$$H_\parallel = \frac{\hbar^2 k_\parallel^2}{2m_0} + \begin{pmatrix} E_c & \frac{\hbar P}{\sqrt{2}m_0}k_+ & 0 & 0 & 0 & -\frac{\hbar P}{\sqrt{6}m_0}k_+ \\ \frac{\hbar P}{\sqrt{2}m_0}k_- & E_v & 0 & 0 & 0 & 0 \\ 0 & 0 & E_v & \frac{\hbar P}{\sqrt{6}m_0}k_- & 0 & 0 \\ 0 & 0 & \frac{\hbar P}{\sqrt{6}m_0}k_+ & E_c & -\frac{\hbar P}{\sqrt{2}m_0}k_- & 0 \\ 0 & 0 & 0 & -\frac{\hbar P}{\sqrt{2}m_0}k_+ & E_v & 0 \\ -\frac{\hbar P}{\sqrt{6}m_0}k_+ & 0 & 0 & 0 & 0 & E_v \end{pmatrix} \tag{A7}$$

where $k_\pm = k_x \pm ik_y$, $k_\parallel^2 = k_x^2 + k_y^2$, $m_0$ is the free electron mass, $E_c$ and $E_v$ are the energies of the conduction and valence band edge and $P = \langle S|p_x|X\rangle = \langle S|p_x|Y\rangle = \langle S|p_x|Z\rangle$ is the momentum matrix element between the conduction and valence band. In SL with layer width $d_1$ and $d_2$, $d_1 + d_2 = d$ values of $E_c$, $E_v$ and $P$ are different in different layers

$$E_c = E_{c1}; E_v = E_{v1}; P = P_1; jd < z < jd + d_1$$

$$E_c = E_{c2}; E_v = E_{v2}; P = P_2; jd - d_2 < z < jd \tag{A8}$$

where $j$ is the number of a SL period. In magnetic field in the growth direction $k_\pm$ can be expressed in Bose operators:

$$k_+ = \frac{\sqrt{2}}{l_B}a; k_- = \frac{\sqrt{2}}{l_B}a^\dagger; k_\parallel^2 = \frac{2}{l_B^2}\left(a^\dagger a + \frac{1}{2}\right) \tag{A9}$$

Solution to Eq.(2) is expanded in functions (4):

$$\Psi(\mathbf{r},z) = \sum_{s=\pm,\alpha} f_{s\alpha}(\mathbf{r})\Xi_{s\alpha}(z) \tag{A10}$$

Substitution of Eq.(A10) in Eq.(A2) leads to the following equations for the coefficients

$$\sum_{s'=\pm,\alpha}\langle s\alpha|H_\parallel|s'\alpha'\rangle = (E - \varepsilon_\alpha)f_{s\alpha} \tag{A11}$$

Off-diagonal matrix elements of $6\times 6$ matrix $\langle s\alpha|H_\parallel|s'\alpha'\rangle$ are nonzero only within groups of states $(+e,+hh,-lh)$ and $(-e,-hh,+lh)$. The reason is that $k_+$ and $k_-$ change $z$-projection of angular momentum $J_z$ by unity and $(-e)$ component with $J_z = 1/2$ can be coupled only with $J_z = 3/2$, i.e. $(+hh)$, and $J_z = -1/2$, i.e. $(-lh)$. As a result, a transposition of rows and columns of $\langle s\alpha|H_\parallel|s'\alpha'\rangle$ reduces it to block diagonal form with two $3\times 3$ equivalent blocks. Then Eq.(A11) is reduced to

$$\begin{pmatrix} \varepsilon_e + \hbar\omega_0\left(a^\dagger a + \frac{1}{2}\right) & \frac{\hbar P_{e,hh}}{m_0 l_B} a & \frac{\hbar P_{e,lh}}{\sqrt{3}m_0 l_B} a \\ \frac{\hbar P_{e,hh}}{m_0 l_B} a^\dagger & \varepsilon_{hh} + \hbar\omega_0\left(a^\dagger a + \frac{1}{2}\right) & 0 \\ \frac{\hbar P_{e,lh}}{\sqrt{3}m_0 l_B} a^\dagger & 0 & \varepsilon_{lh} + \hbar\omega_0\left(a^\dagger a + \frac{1}{2}\right) \end{pmatrix} \begin{pmatrix} f_{+e} \\ f_{+hh} \\ f_{-lh} \end{pmatrix} = E \begin{pmatrix} f_{+e} \\ f_{+hh} \\ f_{-lh} \end{pmatrix} \quad (A12)$$

where $\omega_0 = \hbar / m_0 l_B^2$ and

$$P_{e,\alpha} = P_1 \int_0^{d_1} \xi_e^*(z) \xi_\alpha(z) dz + P_2 \int_{d_1}^{d} \xi_e^*(z) \xi_\alpha(z) dz \quad (A13)$$

Typically $\hbar\omega_0 \ll \frac{\hbar^2 |P|^2}{m_0^2 l_B^2}$ and the resulting equation for energy is:

Equations similar to Eq.(A12) were applied to InAs/GaSb superlattices[23] and 2D hole gas[24-25]. Solution to Eq.(A12) is $f_{+e}(\mathbf{r}) = a_e v_{n-1}(\mathbf{r})$, $f_{+hh}(\mathbf{r}) = a_{hh} v_n(\mathbf{r})$, $f_{-lh}(\mathbf{r}) = a_{lh} v_n(\mathbf{r})$ where $v_n(\mathbf{r})$ is oscillator function of the $n$th level.

$$(\varepsilon_e - E)(\varepsilon_{hh} - E)(\varepsilon_{lh} - E) - (\varepsilon_{hh} - E)\frac{\hbar^2 |P_{e,lh}|^2}{3m_0^2 l_B^2} n - (\varepsilon_{lh} - E)\frac{\hbar^2 |P_{e,hh}|^2}{m_0^2 l_B^2} n = 0 \quad (A14)$$

Sometimes an estimate of energy levels is made with help of the equation for spectrum without magnetic field by replacement of $\frac{\hbar^2 k_\parallel^2}{2m_0}$ with its eigenvalue in magnetic field $(\hbar^2/l_B^2)(n+1/2)$ [16, 26]. The difference compared to the exact result (A14) is erroneous $1/2$. The origin of the error is that off-diagonal elements in Hamiltonian (A7) commute without magnetic field and don't commute in magnetic field. The analytic expression for solution to Eq.(A14) is rather cumbersome. In our experiment $\varepsilon_e - \varepsilon_{hh} \ll \varepsilon_{hh} - \varepsilon_{lh}$. Then in the relevant energy range the cubic equation is reduced to the quadratic one and the solution is:

$$E_n = \frac{\varepsilon_e + \varepsilon_{hh}}{2} \pm \sqrt{\frac{(\varepsilon_e - \varepsilon_{hh})^2}{4} + \frac{\hbar^2 |P_{e,hh}|^2}{m_0^2 l_B^2} n} \quad (A15)$$

Taking zero energy at $\varepsilon_{hh}$, $E_g = \varepsilon_e - \varepsilon_{hh}$ and introducing Fermi velocity as $v_F = \frac{|P_{hh,e}|}{\sqrt{2}m_0}$, we obtain:

$$E_n = \frac{E_g}{2} \pm \sqrt{\frac{E_g^2}{4} + n\frac{2\hbar^2 v_F^2}{l_B^2}} \quad (A16)$$